\documentclass[a4paper,fleqn,usenatbib]{mnras}  
\usepackage{newtxtext,newtxmath}

\usepackage[T1]{fontenc}
\usepackage{ae,aecompl}

\usepackage{graphicx}	
\usepackage{amsmath}	
\usepackage{amssymb}	
\usepackage{indentfirst}
\usepackage{hhline}
\usepackage{adjustbox}
\usepackage{longtable}
\usepackage{multirow}
\usepackage{multicol,tabularx,capt-of}

\title[Accretion of Small Solids]{Simulations of Small Solid Accretion onto Planetesimals  in the Presence of Gas}

\author[A. G. Hughes \& A. C. Boley]{
A. G. Hughes,$^{1}$\thanks{E-mail: ahughes@phas.ubc.ca}
A.C. Boley,$^{1}$
\\
$^{1}$Department of Physics and Astronomy, University of British Columbia, 2329 West Mall, Vancouver V6T 1Z4, Canada
}

\date{Accepted XXX. Received YYY; in original form ZZZ}

\begin{document}
\label{firstpage}
\pagerange{\pageref{firstpage}--\pageref{lastpage}}
\maketitle

\begin{abstract}
The growth and migration of planetesimals in a young protoplanetary disc are fundamental to planet formation. In all models of early growth, there are several processes that can inhibit grains from reaching larger sizes. Nevertheless, observations suggest that growth of planetesimals must be rapid.  If a small number of 100 km sized planetesimals do manage to form in the disc, then gas drag effects could enable them to efficiently accrete small solids from beyond their gravitationally focused cross-section. This gas drag-enhanced accretion can allow planetesimals to grow at rapid rates, in principle.  We present self-consistent hydrodynamics simulations with direct particle integration and gas drag coupling to estimate the rate of planetesimal growth due to pebble accretion. Wind tunnel simulations are used to explore a range of particle sizes and disc conditions.  We also explore analytic estimates of planetesimal growth and numerically integrate planetesimal drift due to the accretion of  small solids.  
Our results show that, for almost every case that we consider, there is a clearly preferred particle size for accretion that depends on the properties of the accreting planetesimal and the local disc conditions.  For solids much smaller than the preferred particle size, accretion rates are significantly reduced as the particles are entrained in the gas and flow around the planetesimal.  Solids much larger than the preferred size accrete at rates consistent with gravitational focusing.
Our analytic estimates for pebble accretion highlight the timescales that are needed for the growth of large objects under different disc conditions and initial planetesimal sizes.  

\end{abstract}


 

\section{Introduction}

The planet formation process is critically dependent on the coagulation of dust grains into larger bodies.  The timescale for significant solid growth is thought to be a few million years, based on the chronologies of meteorites \citep{Vill2009}, the fraction of inferred protoplanetary discs relative to cluster ages \citep{Mamajek2009}, and the incidence of gas giant planets \citep{Cumming2008}.  The latter provides a constraint on core-nucleated instability \citep{Pollack1996} as a dominant formation model for giant planets, which requires the formation of a large solid core before the gaseous disc dissipates.  Furthermore, morphological features in discs such as HL Tau \citep{ALMA} and TW Hydrae  \citep{andrews_twhyd} may indicate the presence of embedded planets, which if correct, would require the rapid growth of large solids in some discs. 

The early stages of planet formation rely on the growth of micron to sub-millimetre grains into planetary sizes.  As grains reach submillimetre and millimetre radii, further growth may be hindered due to bouncing at low collisional speeds and fragmentation at higher collisional speeds \citep{BW2008, Testi2014}. 
Regions of high solid density could potentially build planets or planetesimals through direct gravitational collapse \citep{Gold1973} if solids become concentrated due to some mechanism such as vertical settling \citep{DAl2001}. However, vertical settling is opposed by turbulent mixing, which can lift grains away from the midplane of the disc \citep{Dub1995}, slowing or preventing significant grain growth.  
Furthermore, gas is expected to orbit at sub-Keplerian speeds due to the negative outward pressure gradient in the disc.  This causes objects that are on Keplerian orbits to face a headwind, resulting in the loss of angular momentum for small solids.  
Without any mitigating factor, centimetre to metre-sized objects would consequently spiral into their stars on short timescales  \citep{Adachi1976, Weid1977}.  
While spiral arms or other pressure perturbations could concentrated solids into localized high-pressure regions \citep{Hag2003}, the formation of such structures without the presence of a planet is not well established. 
Altogether radial drift, turbulence, bouncing, and fragmentation hinder the formation and growth of planetesimals and present a major problem for planet formation. 

 A possible solution to this crisis in planet formation is the streaming instability \citep{YG2005, Joh2006}, which relies on solid-gas feedback changing the flow of the gas in the disc.
 If enough solids are concentrated into local regions due to, e.g., local pressure maxima produced by some turbulence, then the solids' collective behaviour can reduce or effectively eliminate drag on the aggregate \citep{YG2005}. This process can allow inward-drifting solids to be collected in the local overdense region until it becomes gravitationally unstable and collapses. This could provide the disc with a population of 100-km sized objects, but cannot explain growth of these planetesimals to planet sizes.
 
Once there is a population of planetesimals, planets can grow directly from collisions between large planetesimals and eventually embryos \citep{Cham2001}, aided by gravitational focusing. Such growth is the basis for terrestrial planet formation as well as giant planet core formation in the nucleated instability theory \citep{Pollack1996}.  Despite many successes, there remain outstanding issues with this model.  For example, growth through pairwise collision can be hindered by bouncing and fragmentation \citep{Lein2011}, and the timescales required to grow giant planet cores often exceeds the observed lifetime of gas in the disc \citep{Teis2013, Fort2007}. 

Planetesimal growth by pebble accretion is an alternative planet formation model to pairwise collision \citep{Lam2012, Joh2015}.  Pebble accretion relies on some mechanism, such as the streaming instability, to rapidly form an initial distribution of rare planetesimals.  Rather than growing planets through collisions between many large planetesimals, pebble accretion posits that these few planetesimals in the disc sweep up the potentially abundant co-orbiting dust and pebbles.   Drag forces between the pebbles and gas in the disc can alter the trajectories of the pebbles as they encounter the planetesimal, causing them to accrete from beyond the planetesimal's gravitational cross-section.  This enables rapid growth of the planetesimal under certain disc conditions and certain planetesimal masses. 

The focus of this work is to investigate the conditions under which pebble accretion is optimized by directly simulating accretion under various conditions, as well as investigating the long-term timescales for growth.  In Section 2, we examine in detail the theoretical framework behind pebble accretion.  Section 3 provides an overview of BOXZY, the hydrodynamics code used in our simulations.  We then present the results of our hydrodynamics simulations of pebble accretion in Section 4, and briefly discuss some of the implications of our results. Finally in Section 5, we estimate the growth and inward drift rates of planetesimals over long timescales. In Section 6 we discuss our main results and their implications, and put our work in the context of previous studies of pebble accretion.

\section{Theory and Methods}

Gas in a protoplanetary disc is expected to orbit at sub-Keplerian speeds due to support from a negative gas pressure gradient.  In the absence of such support, solids will orbit at Keplerian speeds.  This velocity difference causes the solids to encounter a headwind with the gas, and alters the orbit of solids depending on size.  Small solids quickly become entrained with the gas flow, while larger solids carry enough momentum that gas drag only affects their motion over very long timescales.

Throughout this paper, we assume a disc model with the following midplane profiles:
\begin{eqnarray} \label{eqn:disc}
\rho &=& \rho_0 ( \frac{r}{au})^{-n};\\\nonumber
T &=& T_0 (\frac{r}{au})^{-m};\\\nonumber
\end{eqnarray}
where $r$ is the stellar separation measured in au, $\rho_0 = 1 \times 10^{-9} \frac{g}{cm^3}$ and $T_0 =$ 300 K are gas density and temperature values at 1 au.  For our model we use the exponents m = 0.5 and n = 2.5. 
  We take the gas to be ideal with $P = \frac{R_g}{\mu} T \rho$ and $\mu=2.3 \rm~g~mol^{-1}$.  

At any location in the disc, the difference between the circular Keplerian orbit and the circular gas motion is given by
\begin{eqnarray} \label{eqn:vrel}
v_{\textrm{rel}} &=& \sqrt{ \frac{GM}{r}} - \sqrt{\frac{GM}{r} + \frac{r}{\rho}\frac{dP}{dr}} \\ \nonumber
  & = & v_K \left( 1- \sqrt{(1-\left(n+m\right) \left(\frac{ c_0 }{v_K} \right)^2  \left( \frac{r_0}{r} \right)^m  } \right)\\
\end{eqnarray}
for circular Keplerian speed $v_K$ at $r$ and temperature and density power law exponents $m$ and $n$, respectively.  The isothermal sound speed at $r_0=1$ au is given by $c_0$.  The gas conditions at various stellar separations are shown in Table \ref{tbl:discprops}, along with the corresponding relative wind speeds.  For our assumptions, the wind speed $v_\textrm{rel}$ stays roughly constant with stellar separation. 

As a solid experiences a relative wind, drag forces couple them to the gas.  The characteristics of this coupling depends on the drag regime, which can be divided into two basic types: 
Epstein and Stokes.  A solid falls into the Epstein regime when its radius is much smaller than the mean free path (mfp) of the gas, and into the Stokes regime (i.e. the fluid limit) when the radius is much larger than the mfp.  The Stokes regime itself has different drag behaviours depending on the Reynolds number of the gas and the Mach speed of the particle through the gas \citep{Whip1972}. Consequently, whether a given particle is in the Epstein or Stokes limit  depends on the local conditions of the disc as well as the properties of the particle.  

Let $\lambda\approx \frac{\bar{m}}{\rho \sigma}$  represent the gas's mfp, in which ${\bar{m}}$ is the typical mass of a gas particle, $\rho$ is the  local gas density, and  $\sigma\sim 2.0 \times 10^{-15} \textrm{cm}^2$ (for H$_2$) is the characteristic cross section of a gas particle \citep{Allen}.  
%
%
%
%
%
The division between particle sizes for the Epstein and Stokes regimes is dependent on the density of the disc, which indirectly leads to a dependence on stellar separation. 
At 1 au, $\rho\sim10^{-9} \frac{g}{cm^{3}}$ and $\lambda \sim 2$ cm.  At 10 au, $\lambda\sim 500$ cm.  For disc radii larger than 1 au, most of the solids that we will consider will be in the Epstein regime.  Interior to 1 au, even submillimetre grains can be in the Stokes fluid limit.  

%

%

In the Epstein limit, the stopping time of the gas is given by
\begin{equation}
\label{eqn:epstop}
\tau_f =\frac{s \rho_s}{\rho v_{\rm th}},
\end{equation}
where $s$ is the solid size, $\rho_s$ is the internal solid density, $v_{\rm th}$ is the mean thermal speed of the gas, and $\rho$ is the gas density.  
The stopping time can be thought of as the e-folding time needed for solid-gas flow coupling without any additional external forces. 
For solids in the Stokes limit at low Reynolds number,
\begin{equation} \label{eqn:ststop}
\tau_f = \frac{2}{9} \frac{s^2 \rho_{s}}{\eta}.
\end{equation}
Here, $\eta$ is the molecular viscosity in the disc.

In either regime, particles with small $\tau_f$ rapidly lose angular momentum to the gas, but because $\tau_f$ is so low, the radial drift is also negligible. 
Large solids have correspondingly large $\tau_f$ and lose angular momentum only over very long timescales.  In between these sizes, when the Stokes number ${\rm St}=\tau_f \Omega \approx 1$ for gas orbital frequency $\Omega$, solids will exhibit motion that is affected by gas drag without being perfectly coupled to the gas.  For these sizes, significant radial drift of the solids can occur.

\begin{table}
\centering
\begin{tabular}{|l||l|l|l|l|}
\hline
$r$ (au) & $\rho_\textrm{g}$ $\frac{g}{cm^3}$ & T($K$) & $v_{\textrm{rel}}$ $\frac{m}{s} $ & s(cm) \\ \hhline{|=||=|=|=|=|}
10                    & $3.2 \times 10^{-12} $       & 95              & 54.8       &0.0005                      \\ \hline
3                     & $6.4 \times 10^{-11}$        & 170             & 54.7       &0.01                      \\ \hline
1                     & $1.0 \times 10^{-9}$         & 300             & 54.7        &0.2                     \\ \hline
0.3                   & $2.0 \times 10^{-8}$         & 550             & 54.6        &3.0                     \\ \hline
0.1                   & $3.1 \times 10^{-7} $        & 950             & 54.6        &50.0                     \\ \hline
\end{tabular}
\caption{\small{Disc and object properties used in the hydrodynamics simulations, following the flared disc profile outlined in Equation (\ref{eqn:disc}) along with predicted best-accreted pebble size for each distance as found using Equation (\ref{eqn:bestpebble}).  While most disc conditions change dramatically at different stellar separations, the relative velocity between the gas and the planetesimal stays roughly constant.}}\label{tbl:discprops}
\end{table}

Particle accretion onto a planetesimal can also be affected by gas-drag effects.   In the absence of any gas, solids will be accreted by a planetesimal if their impact radii are within the planetesimal's gravitationally focused cross section 
\begin{equation} \label{eqn:gravcs}
\sigma_{\rm grav} = \pi R^2 \left(1 + \Big( \frac{v_{\textrm{esc}}}{v_{\textrm{rel}}} \Big)^2\right) 
\end{equation}
for planetesimal radius $R$. 
The presence of gas can increase or even decrease the effective accretion cross section of the planetesimal, depending on the gas conditions and the particle size.  
%
%
To see this, consider the solid stopping distance
\begin{equation} \label{eqn:dstop}
d_{\textrm{stop}} = v_{\textrm{rel}} \tau_f. 
\end{equation}
Accretion is optimized when the stopping distance is comparable to the Bondi radius of the object onto which the pebbles are accreted.  The Bondi radius is defined here as
\begin{equation}
\label{eqn:bondir}
\textrm{R}_\textrm{b} = \frac{G M }{v_\textrm{rel}^2},\\
\end{equation}
\\
which is the size at which the kinetic energy of the incoming pebbles becomes approximately equal to the gravitational potential due to the planetesimal. If a significant fraction of a pebble's kinetic energy is dissipated while it traverses the volume defined by the Bondi radius, then the pebble can be accreted onto the planetesimal.  When $d_\textrm{stop} \sim \textrm{R}_\textrm{b}$ the pebble is not perfectly coupled to the gas but still experiences drag during its encounter.  We estimate the most efficient accretion size in the Epstein regime by equating $d_\textrm{stop}$ with $\textrm{R}_\textrm{b}$, yielding
\begin{equation} \label{eqn:bestpebble}
s = \frac{\rho G M}{\rho_s v_{\textrm{rel}^3}} \sqrt{\frac{8 R_g T}{\pi\mu}},
\end{equation}
where $\mu = 2.3 \frac{g}{\textrm{mol}}$ is the gas's mean weight.

In practice, the Bondi radius cannot exceed the Hill radius of the planetesimal.  This condition is not relevant in the simulations explored here, as the planetesimal is never large enough to approach this limit.  Table \ref{tbl:discprops} shows the ideal particle size for each stellar separation, given by Equation (\ref{eqn:bestpebble}).

We investigate the accretion rates of a range of pebble sizes at the stellar separations listed in Table \ref{tbl:discprops}.  In order to model planetesimal growth at different distances, we use a Cartesian grid with particles streaming past a centred planetesimal.  The movement of gas is modeled by a wind tunnel, where the speed of oncoming gas is set by the relative velocity between the planetesimal and sub-Keplerian orbiting gas.  Particles are introduced onto the grid at velocities equal to the relative gas speeds, adapting to friction with the gas.  To recover instantaneous accretion rates for a range of disc conditions and pebble sizes, simulations are run over short timescales until the accretion rates are stabilized.  Over such short timescales, we do not consider inward radial drift of the planetesimal or pebbles.\\

\section{Boxzy}

Simulations are run using Boxzy Hydro \citep{Boley2013}, a Cartesian, finite volume hydrodynamics code.  Boxzy includes a detailed equation of state with rotational and vibrational $H_2$ states, self-gravity (not used here), direct particle integration, and drag terms with feedback on the gas.  Boxzy solves the hydrodynamics continuity equations in the following form: 
\begin{align}
\frac{\partial \rho}{\partial t} + \vec{\bigtriangledown} \cdot (\vec{v} \rho) &= 0 \label{eqn:masscont} \tag*{(Mass Continuity), } \\ \frac{\partial \rho \vec{v}}{\partial t} + \vec{\bigtriangledown} (\vec{v} \rho \vec{v}) &= - \vec{\bigtriangledown} P - \rho \vec{\bigtriangledown} \Phi \label{eqn:momtm} \tag*{(Momentum Equation),} \\
\frac{\partial E}{\partial t} + \vec{\bigtriangledown} \cdot (\vec{v} (E + P)) &= - \rho \vec{v} \cdot \vec{\bigtriangledown} \Phi \label{eqn:ener} \tag*{(Energy Equation), }
\end{align}
where $E =\frac{1}{2} \rho |v|^2 + \epsilon$, for a gas density $\rho$, velocity v, and internal energy density $\epsilon$. 

The code uses a Reimann solver for shock capturing.  The hydrodynamic state is advanced by first solving the hydrodynamics equations using the current state to predict the state at the end of the time step, and then uses that prediction to update the final state.  The inclusion of gravity is solved using a kick-drift-kick method during sourcing.

The gas is evolved adiabatically, comprised of a mixture of $H_2$ (mass fraction 0.73), He (mass fraction 0.25), and metals (mass fraction 0.02) such that $\mu = 2.3 \frac{g}{\textrm{mol}}$.  The rotational and vibrational modes of $H_2$ are included as described in \citep{Boley2007} for a fixed 3:1 ortho-parahydrogen mixture.  

Boxzy can be run in 2D, 3D, or 2D-3D.  We use the latter configuration in this study. The 2D-3D simulations use a cylindrical cross-section to capture 3D effects on a 2D grid, creating a wind tunnel around the planetesimal, set in the reference frame of the planetesimal.  The ``wind'' is the relative velocity between the gas/small solids and the planetesimal. 

\begin{figure}
\centering
\includegraphics[scale=0.38]{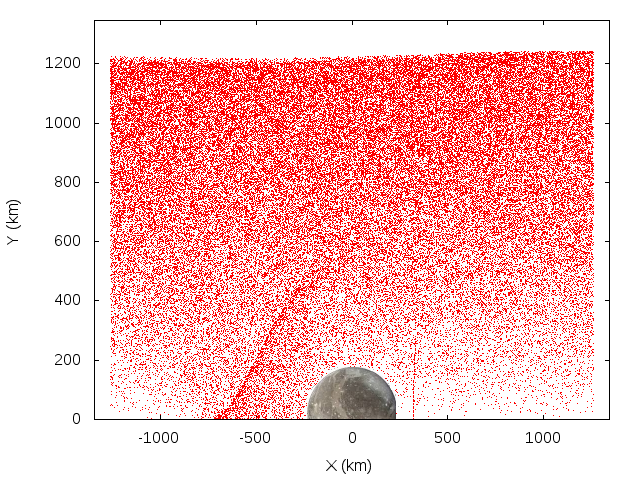}
\caption{\small{Distribution of particles on the grid - the higher population at larger y-values takes account of the cylindrical geometry. A clear and physical wake has formed to the left of the planetesimal due to the particle and gas flow.  The stack of pebbles approaching the planetesimal from the right are the 60 trace particles used to follow trajectories at different impact radii.  Finally, the slight deviation seen in all particles near the top of the plot is a result of gravitational focusing.}}\label{fig:wake}
\end{figure}

Our simulations use a grid of 256 $\times$ 128 cells, where each cell has the dimensions of 10 km $\times$ 10 km or 15 km $\times$ 15 km.  The different size cells are chosen such that the planetesimal radius is always resolved by 10 cells.  We use a 100 km and a 150 km sized planetesimal, horizontally centred on the grid as shown in Figure \ref{fig:wake}.  In the 2D-3D setup, we exploit azimuthal symmetry about the x-axis; thus, the y coordinate is the cylindrical radial distance from the x-axis. 

\begin{figure}
\centering
\includegraphics[scale=0.15]{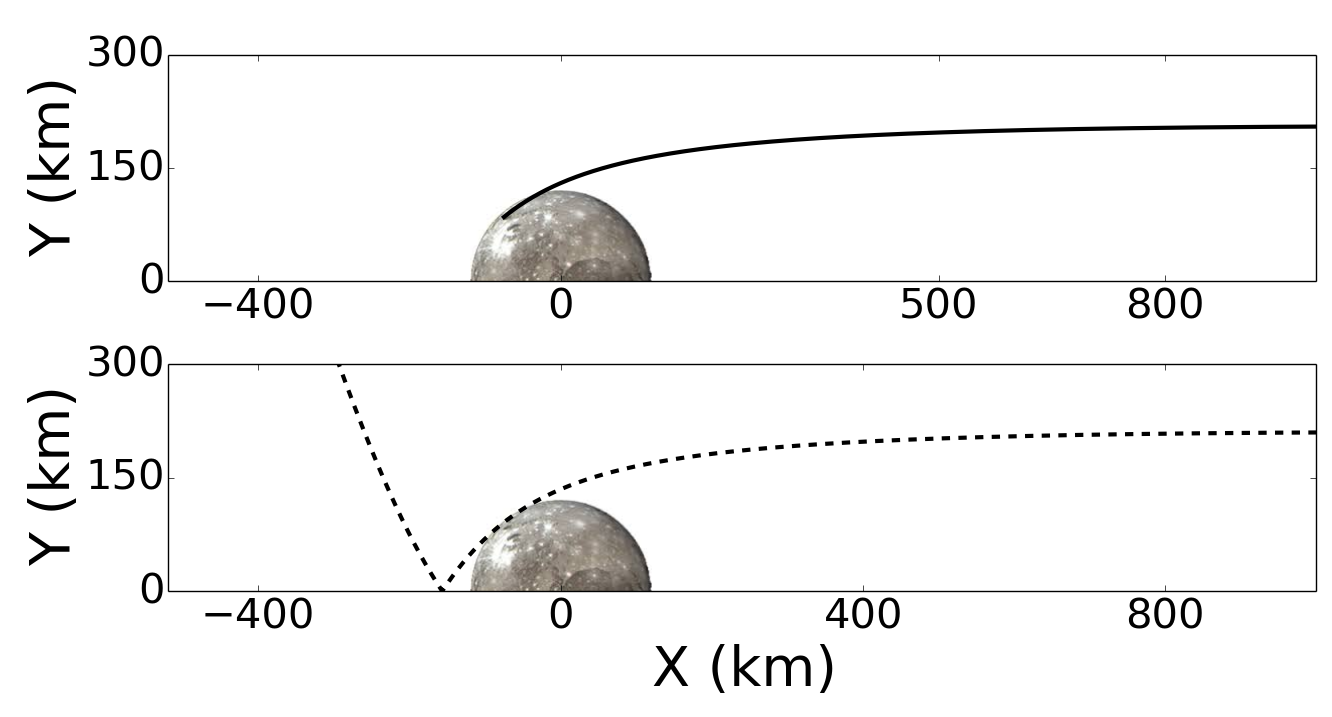}
\caption{\small{Two traced 0.1 cm particles at 10 au, separated by a distance dy, 10 km.  The particle on top is accreted, whereas the bottom one flows past the planetesimal and is reflected upwards, taking account of particles that would come in from the negative half of the y-axis.}}\label{fig:vertstack}
\end{figure}

Particles are coupled to the gas using a cloud-in-cell approximation and a kick-drift-kick integration algorithm.  The gas drag and gravity kicks are applied while sourcing the gas.  All particles that hit the x-axis are reflected back upwards with the same velocity components (but opposite sign in the $y$ direction), taking account of solids that would enter the grid from below the planetesimal. This effect can be seen in the bottom panel of Figure \ref{fig:vertstack}.

Solids in the simulation are modeled by ``superparticles''; i.e., each particle on the grid represents a swarm of real particles.  The grid  can be thought of as a wedge of a cylindrical volume surrounding the planetesimal, such that the placement of the particles must take this geometry into account to ensure superparticles are isotropically distributed in 3D space. This requires that more particles are placed at higher $y$-values. The placement of a given particle at a location $Y$ along the $y$ coordinate is determined using 
\begin{equation} \label{eqn:yplacement}
Y = (n_y) \times (dy) \times \sqrt{U},
\end{equation}
where $n_y$ is the number of cells in the y-direction, $dy$ is the length of each cell, and $U$ is some uniform random variate between 0 and 1.  Figure \ref{fig:wake} shows the location of particles for one snapshot of one of the wind tunnel simulations.  The larger number of particles at high $y$ reflect represent an isotropic placement in the 3D volume.  


Of the 100,000 particles, Boxzy records in detail the positions and velocities of 60 particles, stacked in the y-direction from 0 to 300 km for simulations with a 100 km planetesimal (450 km for the 150 km planetesimal), allowing us to track how particles are accreted or become deflected around the planetesimal (Figure \ref{fig:vertstack}).
The traced particles are apparent in Figure \ref{fig:wake} (for the given snapshot) as a line just in front of the planetesimal.

Whenever any particle (traced or not) makes contact with the planetesimal, it is accreted, recorded, and reset to enter along with the incoming wind according to Equation (\ref{eqn:yplacement}).  Because the particle distribution takes into account the cylindrical geometry, the accretion rate is directly related to the number of particles striking the planetesimal.  The particle-in-cell algorithm effectively spreads a particle over the nearest 4 cells (in 2D-3D mode), with a weighting related to the distance to a given cell centre.  If any part of the cloud contacts the planetesimal, the entire superparticle is accreted.  As such, this slightly enhances the physical size of the planetesimal, which is taken into account when we compare our results with analytic expectations for a limiting case.

We track the total number of accreted particles per unit time, $\frac{\Delta N}{\Delta t}$, during the simulation.  This enables us to calculate the corresponding accretion rate, $\dot{M} $.  Each superparticle represents a fixed amount of mass in solids given by
\begin{equation}\label{eqn:supermass}
M_s = \frac{\rho f}{N_{\textrm{part}}} \pi {L_y}^2 L_x,
\end{equation}
where $\rho$ is the unperturbed gas density, $f$ is the dust-to-gas ratio, assumed to be 0.01 for the simulations, $L_y$ is the simulation length in the y-direction (radial), and $L_x$ is the length along the symmetry axis.   The accretion rate is then simply $\dot{M} = \frac{\Delta N}{\Delta t} M_s$.  We can in principle choose different dust-to-gas ratios provided that they do not alter the particle trajectories.  For the simulations, we set a low $f$ so that particle feedback on the gas flow is negligible.

After the accretion rate for a given simulation has been determined, that rate can be used to estimate the time needed to increase the planetesimal mass by a factor of e.  In this estimate, we assume a fixed accretion rate and assume that all solids have the same size, i.e.,  
\begin{equation} \label{eqn:massgrowth}
t_{\textrm{M}} = \frac{M_{\textrm{i}}}{\dot{M}},
\end{equation}
where $M_{\textrm{i}}$ is the initial mass of the planetesimal and $\dot{M}$ is the given accretion rate.

\section{Simulation Results}

We ran 132 simulations using the 2D-3D Boxzy configuration, exploring disc locations between 0.1 au and 10 au and particle sizes ranging from 1 $\mu$m to 100 cm (depending on the disc location).

Different planetesimal radii and densities were also used, in which planetesimals with an internal density of $1.5 \frac{g}{cm^3}$ were run for a planetesimal size $R=100$ km and planetesimals with an internal density of $3 \frac{g}{cm^3}$ were run with $R=100$ km and $R=150$ km.
These sizes are used because they represent the approximate upper end of the initial planetesimal distribution  produced by the streaming instability in a solar-like protoplanetary disc  \citep{Joh2015}. 
  In all cases, small particles are assumed to have an internal density of $3 \frac{g}{cm^3}$ for drag calculations.

\begin{figure*}
\centering
\includegraphics[scale=0.37]{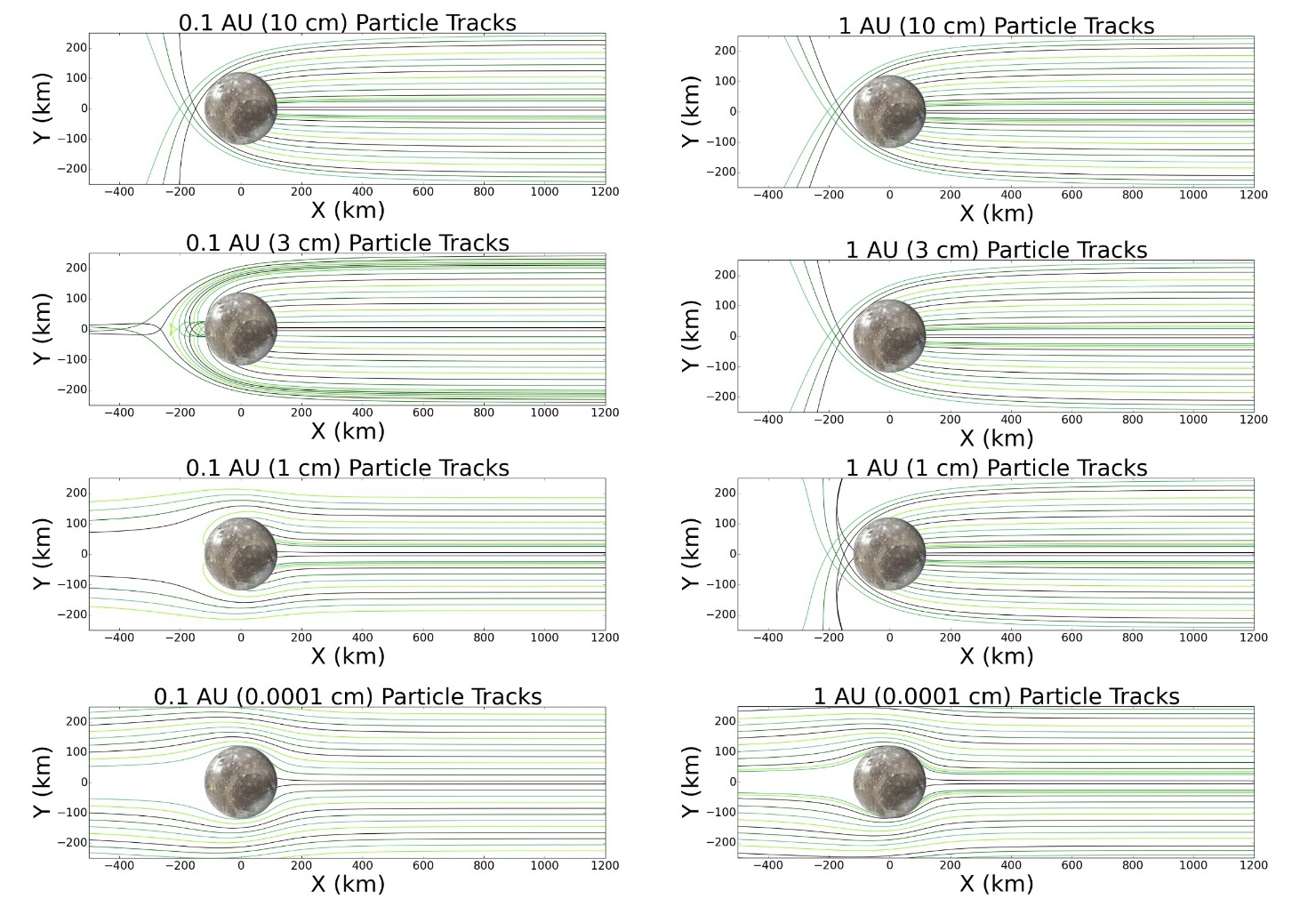}
\caption{\small{Traced particle tracks at 0.1 au (left panel) and 1 au (right panel) for various different particle sizes, showing their movement as they are accreted onto or deflected around the planetesimal.  The largest pebble size, 10 cm, comes in from beyond the geometric radius and is efficiently accreted. The most dramatic looping pattern is seen with the 3 cm particles at 0.1 au, which accrete onto the planetesimal even after they have moved past it significantly. At smaller sizes, accretion becomes much less efficient.  The 0.0001 cm particles mostly stream past the planetesimal, and are accreted from a smaller area than even its geometric cross-section.} }\label{fig:sdbyside}
\end{figure*}

The curves in Figure \ref{fig:sdbyside} show the trajectories of solids as they encounter the planetesimal with different impact radii, disc locations, and solid sizes.  For example, at 1 au the gas density is $\rho = 1 \times 10^{-9} \frac{g}{cm^3} $, and the most efficiently accreted solid size is $\sim$0.3 cm.   
Solids much larger than this follow gravitational focusing, while solids much smaller simply flow around the planetesimal.

\begin{figure*}
\centering
\includegraphics[scale=0.34]{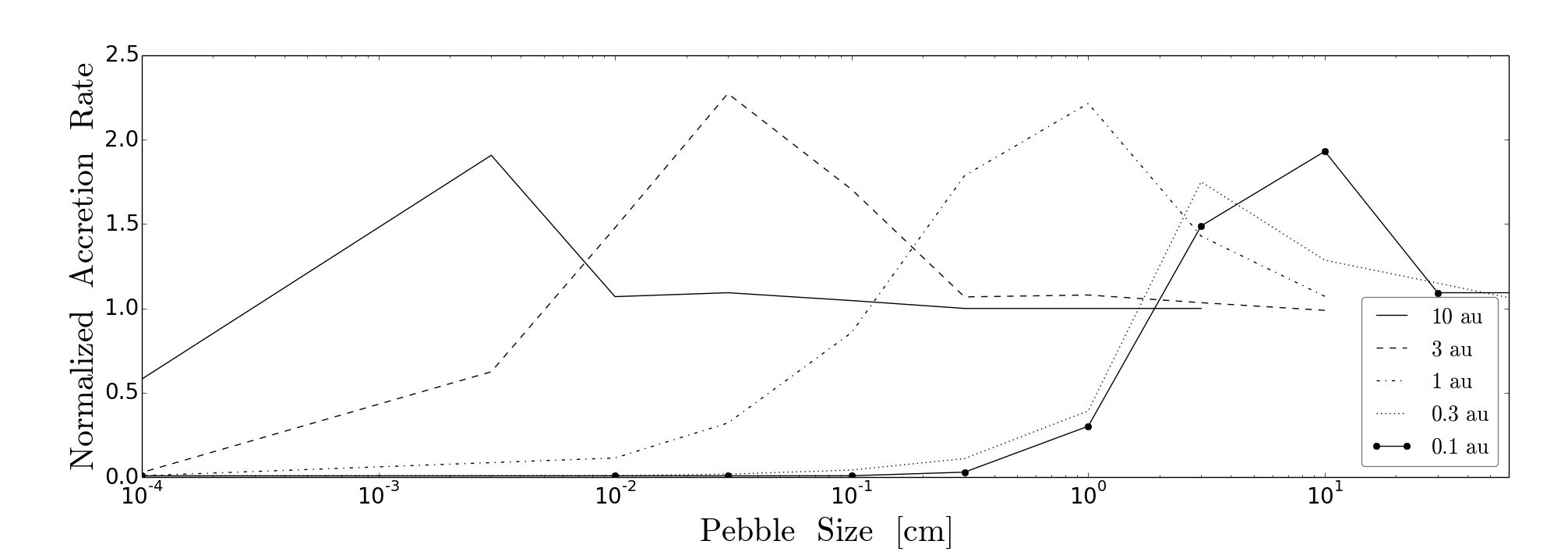}
\caption{Accretion rates for each distance are plotted against pebble sizes for the $\textrm{M}(100)_{3.0}$ simulation. The rates are normalized to those expected from gravitational focusing only (Equation \ref{eqn:acc_grav_rate}). The pebble sizes are plotted on a logarithmic scale.  Note that each distance has a preferred pebble size, with farther distances favouring smaller pebbles and closer distances favouring larger ones.}\label{fig:final_accr}
\end{figure*}

Accretion rates for all simulations are shown in Table \ref{tbl:bigtable}. There is a clear preferred solid size for most situations, but because we only have coarse sampling, the absolute peak of the accretion rate may be missed.  Nevertheless, the overall trends are apparent. Accretion is strongly suppressed for solids much smaller than the preferred size, which leads to accretion at rates even lower than what is predicted from the geometric cross-section alone.  For particles much larger than the preferred size, the accretion rates tend toward those expected due to gravitational focusing.  To help visualize the results, we show  in Figure \ref{fig:final_accr} the normalized accretion rates for the $\textrm{M}(100)_{3.0}$ simulations as a function of particle size.   The normalization is set by the gravitational focusing limit, with a corresponding accretion rate   

\begin{equation}\label{eqn:acc_grav_rate}
\dot{M}_G = f \rho \pi R^2 v_{\textrm{rel}} F_g,
\end{equation}
in which
\begin{equation}\label{eqn:gf}
F_g = 1 + \frac{8}{3} \pi G \frac{R^2 \rho_s}{(v_{\textrm{rel}})^2}
\end{equation}
for a given planetesimal size $R$ and planetesimal density $\rho_s$. 
For each planetesimal's location, size, and density, the corresponding $\dot{M}_G$ can be compared with the accretion rates in Table \ref{tbl:bigtable}.   Figure \ref{fig:final_accr} shows that the simulations do approach the gravitational focusing accretion rates whenever the particle sizes are much larger than the preferred accretion size, which should be the case.   For broader comparison with Table \ref{tbl:bigtable}, we expect the accretion rates in the gravitational focusing limit to be  $\dot{\textrm{M}}(100)_{3.0} = 430 \frac{\textrm{kg}}{\textrm{s}}$, $4.3 \times 10^{7} \frac{\textrm{kg}}{\textrm{s}}$ at 10 au and 0.1 au, respectively; and  $\dot{\textrm{M}}(150)_{3.0} = 2000 \frac{\textrm{kg}}{\textrm{s}}$, $2.0 \times 10^{8} \frac{\textrm{kg}}{\textrm{s}}$ at 10 au and 0.1 au, respectively. 

As noted, Boxzy uses the cloud-in-cell method, in which individual particles are spread over the nearest four surrounding cells.  If any one of the four cells includes the planetesimal, the superparticle is assumed to be accreted.  This gives the planetesimal an effective radius of 11 cells instead of 10 (an extra 10 km or 15 km for the 100-km or 150 km planetesimal, respectively).  This size difference is taken into account in the above values.  The agreement between our largest particle size accretion rates and the expected accretion rates from gravitational focusing is taken as evidence that the simulation box is large enough to properly capture gravitational focusing effects and accretion trajectories.  

\begin{table*}
\centering
\caption{\small{Accretion rates for a 100 km and a 150 km planetesimal at different distances ($r$) ranging from 0.1 au to 10 au. The 100 km planetesimal results are shown with planetesimal densities $1.5 \frac{g}{cm^3}$ as $\dot{M}(100)_{1.5}$ in column 3, $3 \frac{g}{cm^3}$ as $\dot{M}(100)_{3.0}$ in column 4, and accretion rates for a 150 km planetesimal (only $3 \frac{g}{cm^3}$) in column 5.  The most efficient accretion rates are bolded whenever they are significantly higher for a certain particle size.  Column 6 shows the ratio of the Bondi radius to the stopping distance for the 100 km planetesimal with an internal density of 3 $\frac{g}{cm^3}$. }}
\begin{adjustbox}{width=0.80\textwidth}
\label{tbl:bigtable}
\begin{tabular}{|
>{}l ||l|l|l|l|l|l|}
\hline
\textbf{$r$(au)} & \textbf{Particle Size (cm)} & {$\dot{M}(100)_{1.5} (\frac{kg}{s}) $}             & \textbf{$\dot{M}(100)_{3.0}  (\frac{kg}{s}) $} &\textbf{$\dot{M}(150)_{3.0} (\frac{kg}{s})$}     &$\frac{R}{d_s}$  \\ \hhline{|=||=|=|=|=|=|}
\multicolumn{1}{|c||}{10}                                                     & 3                                                   & 250                                                                                                 & 430                  &   1900                          &1.3$\times 10^{-4}$                                                                                                            \\ \hline
                                                       & 1                                                   & 250                                                                                                 & 430           & 1900                      &      4.1$\times 10^{-4}$                                                                                                    \\ \hline
                                                       & 0.3                                                 & 250                                                                                                 & 430         & 1900  & 1.3$\times 10^{-3}$                                                                                                                                \\ \hline
                                                       & 0.1                                                 & 250                                                                                                 & 450                      & 2000       & 4.1 $\times 10^{-3}$                                                                                                               \\ \hline
                                                       & 0.03                                                & 250                                                                                                 & 470        &3500            & 1.3$\times 10^{-2}$                                                                                                                       \\ \hline
                                                       & 0.01                                                & 250                                                                                & 460  & 6600                          & 4.1 $\times 10^{-2}$                                                                                                                                      \\ \hline
                                                       & 0.003                                               & 240                                                                                                 & \textbf{820}       &\textbf{8400}                                                                                     &0.13                                                                             \\ \hline
                                            & 0.0001                                              & 120                                                                                                 & 250           & 940                    &4.1                                                                                                                                  \\ \hhline{|=||=|=|=|=|=|}
\textbf{$r$ (au)} & \textbf{Particle Size (cm)} & \textbf{$\dot{M}(100)_{1.5} (\frac{kg}{s}) $}             & \textbf{$\dot{M}(100)_{3.0}  (\frac{kg}{s}) $}   & \textbf{$\dot{M}(150)_{3.0} (\times 10^4 \frac{kg}{s})$}     &$\frac{R}{d_s}$ \\ \hhline{|=||=|=|=|=|=|}
\multicolumn{1}{|c||}{3}                                                      & 10                                                  & 5100                                               & 8700     & 3.9                  & 0.001                                                                                                                  \\ \hline
& 3       & 5200                                                                                                & 9100 & 4.1    & 0.0036                                                                                                                                      \\ \hline
                                                       & 1                                                   & 5100                                                                                                & 9500 & 6.8  &0.011                                                                                                                                         \\ \hline
                                                       & 0.3                                                 & 5000                       & 9400   & 13            &0.036                                                                                                                             \\ \hline
                                                       & 0.1                                                 & 4800                                                                                                & 15000    & \textbf{17}        &0.11                                                                                                                             \\ \hline
                                                       & 0.03                                                & \textbf{6600}                                                                                                & \textbf{20000} & 12                 &0.36                                                                                                                        \\ \hline
                                                       & 0.01                                                & 5600                                                                                                & 13000 & 5.7             &1.1                                                                                                                          \\ \hline
                                                       & 0.003                                               & 2500                                                                                                & 5500   & $0.44$               &3.6                                                                                                                       \\ \hline
                                                       & 0.0001                                              & 150                                                                                                & 250 & 0.084          &110                                                                                                                              \\ \hhline{|=||=|=|=|=|=|}
\textbf{$r$ (au)}                      & \textbf{Particle Size (cm)} & \textbf{$\dot{M}(100)_{1.5} (\times 10^4 \frac{kg}{s}) $} & \textbf{$\dot{M}(100)_{3.0} (\times 10^4 \frac{kg}{s}) $} & \textbf{$\dot{M}(150)_{3.0} (\times 10^5 \frac{kg}{s}$)}  &$\frac{R}{d}$\\ \hhline{|=||=|=|=|=|=|}
\multicolumn{1}{|c||}{1}                                                      & 10                                                  & 7.9                                                                                                & 15                  &17                  &0.02                                                                                                      \\ \hline
                                                       & 3                                                   & 7.7                                                                                                & 20                  &\textbf{26} &0.067                                                                                                                                        \\ \hline
                                                       & 1                                                   & 8.8                                                                                                & \textbf{31}                 &23   &0.2                                                                                                                                      \\ \hline
                                                       & 0.3                                                 & \textbf{10}                                                                                                & 25 &12            &0.67                                                                                                                             \\ \hline
                                                       & 0.1                                                 & 5.7                                                                                                & 12 &4.9    &2.0                                                                                                                                  \\ \hline
                                                       & 0.03                                                & 2.1                                                                                                & 4.5 &1.6         &6.7                                                                                                                          \\ \hline
                                                       & 0.01                                                & 0.79                                                                                               & 1.6     &0.61           &20                                                                                                                       \\ \hline
                                                       & 0.0001                                              & 0.13                                                                                               & 0.13      & 0.028               &2000                                                                                                \\ \hhline{|=||=|=|=|=|=|}
\textbf{$r$ (au)}                      & \textbf{Particle Size (cm)} & \textbf{$\dot{M}(100)_{1.5} (\times 10^4 \frac{kg}{s}) $} & \textbf{$\dot{M}(100)_{3.0} (\times 10^4 \frac{kg}{s}) $} & \textbf{$\dot{M}(150)_{3.0} (\times 10^5 \frac{kg}{s}$)}   &$\frac{R}{d}$ \\ \hhline{|=||=|=|=|=|=|}
\multicolumn{1}{|c||}{0.3}                                                    & 100                                                 & 160                                                                                                 & 280         &120       &1.0$\times 10^{-4}$                                                                                                                           \\ \hline
                                                       & 10                                                  & 160                                                                                                 & 360     &\textbf{500}     &0.01                                                                                                                                 \\ \hline
                                                       & 3                                                   & \textbf{200}                                                                                                 & \textbf{490}     &230        &0.11                                                                                                                              \\ \hline
                                                       & 1                                                   & 54                                                                                                & 110 &42         &1.0                                                                                                                              \\ \hline
                                                       & 0.3                                                 & 15                                                                                                & 31 &11               &11.0                                                                                                                           \\ \hline
                                                       & 0.1                                                 & 6.0                                                                                                & 12    &4.2            &100                                                                                                                        \\ \hline
                                                       & 0.03                                                & 3.2                                                                                                & 5.2 &1.4         &1100                                                                                                                                \\ \hline
                                                       & 0.01                                                & 2.6                                                                                                & 2.9    &0.68            &1.0$\times 10^{4}$                                                                                                                         \\ \hline
                                                       & 0.0001                                              & 2.5                                                                                               & 2.5   &0.42           &1.0$\times 10^{8}$                                                                                                                           \\ \hhline{|=||=|=|=|=|=|}
\textbf{$r$ (au)}                      & \textbf{Particle Size (cm)} & \textbf{$\dot{M}(100)_{1.5} (\times 10^5 \frac{kg}{s}) $} & \textbf{$\dot{M}(100)_{3.0} (\times 10^5 \frac{kg}{s}) $}  & \textbf{$\dot{M}(150)_{3.0} (\times 10^6 \frac{kg}{s}$)}  & $\frac{R}{d}$\\ \hhline{|=||=|=|=|=|=|}
\multicolumn{1}{|c||}{0.1}                                                    & 100                                                 & 260                                                                                                 & 470        & 250                                 &1.4$\times 10^{-4}$                                                                                                 \\ \hline
                                                       & 30                                                  & 250                                                                                                 & 470 & 560 &1.6 $\times 10^{-3}$                                                                                                                                          \\ \hline
                                                       & 10                                                  & 240                                                                                                 & \textbf{830}                               &\textbf{790}  &0.014                                                                                                                                     \\ \hline
                                                       & 3                                                   & \textbf{290}                                                                                                 & 640  &280   &0.16                                                                                                                                     \\ \hline
                                                       & 1                                                   & 63                                                                                                & 130 &46  &1.4                                                                                                                                  \\ \hline
                                                       & 0.3                                                 & 8.1                                                                                                & 13 &4.7     &16.0                                                                                                                                    \\ \hline
                                                       & 0.1                                                 & 3.8                                                                                                & 3.9 &6.2               &140                                                                                                                          \\ \hline
                                                       & 0.03                                                & 3.7                                                                                                & 3.7          &0.64     &1600                                                                                                                         \\ \hline
                                                       & 0.01                                                & 3.6                                                                                                & 3.7  & 0.62                  &1.4$\times 10^4$                                                                                                                   \\ \hline
                                                       & 0.0001                                              & 3.6                                                                                                & 3.6      & 0.62      &1.4$\times10^{8}$                                                                                                                             \\ \hline
\end{tabular}
\end{adjustbox}
\end{table*}



 At 10 au, gas-enhanced accretion is only seen for particles $\lesssim 100 ~\mu m$ and only for the $3 \frac{g}{cm^3}$ planetesimals, which is consistent with expectations from Equation (\ref{eqn:bestpebble}).   
 The lowest mass planetesimal (density $1.5 \frac{g}{cm^3}$) at 10 au accretes all but micrometre sizes at roughly the same rate.  This implies that accretion onto such planetesimals would reflect the size distribution of particles in the disc.  At closer distances, accretion rates are optimized for particles of a few centimetres in size, consistent with expectations from Table \ref{tbl:discprops}. In most cases, the gas-drag enhanced accretion rates are larger than the expected gravitational focusing rates by a factor of two to four.  The largest enhancements are seen with the $R=150$ km planetesimal.
Particles much smaller than the most efficient size have considerably lower accretion rates, even a few orders of magnitude lower than estimates from gravitational focusing. This indicates that particles must undergo growth beyond micrometre regimes before being effectively accreted onto a growing planetesimal in the presence of gas. 
 


 Accretion rates at 0.1 au exceed those at 10 au by up to 5 orders of magnitude, which is due to decreasing gas density with stellar separation (Equation \ref{eqn:gravcs}).  This highlights the noted difficulty in planetesimal formation at large stellar separations and is emphasized further by extrapolating our accretion rates to estimate mass growth. 
 
\subsection{Extrapolated Mass Growth Estimates} 

Using Equation \ref{eqn:massgrowth}, we calculate the mass growth timescales $t_{\rm m}$ for all three planetesimal cases: R = 100 km with $\rho = 1.5 \frac{g}{cm^3}$ and $\rho = 3 \frac{g}{cm^3}$, and R = 150 km with $\rho = 3 \frac{g}{cm^3}$.   For each orbital distance and planetesimal, we calculate the mass growth timescale by selecting the highest $\dot{M}$ values found in our simulations.  We assume a dust-to-gas ratio of $f=0.1$ in these values.   We find that at 10 au, the $\textrm{M}(100)_{3.0}$ planetesimal's mass would increase by a factor $e$ on a $t_\textrm{m} = 5.0 \times 10^{7}$ year timescale, while at 0.3 au that decreases to 8200 years, and again down to 480 years at 0.1 au. The timescales are roughly a factor of 3 shorter for the $\textrm{M}(150)_{3.0}$ planetesimal, and a factor of 1.5 larger for the $\textrm{M}(100)_{1.5}$ planetesimal.  

At large stellar separations,  the accretion timescales exceed the lifetime of gas in the disc. At distances  $r\lesssim 3$ au, however, significant growth can occur.  These results are for the given conditions of our model disc and the selected planetesimals.  Moreover, they do not take into account the runaway process inherent in either mass growth mechanism (gravitational focusing or Bondi accretion). 
To explore a wider range of conditions, we next look at analytic estimates for growth.

\section{Idealized Growth and Drift Models}

Our simulations are run over short timescales, providing only instantaneous accretion rates.  
While these rates can be used to gain insight on the long-term growth of the planetesimal, they do not take into account the changing mass of the planetesimal or, consequently, the evolving accretion rates due to increased gravitational or drag focusing.  
Analytics can take such changes into account, at least under simplifying assumptions. 
Numerical integration can also address secondary effects, such as the migration of the planetesimal itself due to mass growth (and drag) from a lower angular momentum population of pebbles. 


\subsection{Idealized mass growth}

The planetesimals studied here will grow in size from the accretion of small solids until the flow of those solids is arrested by an external factor (e.g., loss of supply, Hill-sphere truncation, etc.).
Under simplifying conditions, we can determine the timescale for which we can expect the planetesimal to have a rapid change in size due to a runaway accretion rate.  
First, consider the mass accretion rate for a planetesimal in the midplane of a disc:
\begin{equation} \label{eqn:mdot_planetesimal}
\dot{M} = f \rho  v_{\textrm{rel}} \pi {R^\prime}^2,
\end{equation}
where $\rho$ is the midplane gas density, $f$ is the midplane dust-to-gas ratio, $ v_{\rm rel}$  is the relative speed between the planetesimal and the small solid population, and $R'$ is the effective accretion radius of the planetesimal.  
Throughout the calculations, we assume that $v_{\rm rel}$ represents a headwind.
If the planetesimal grows with an approximately constant internal density $\rho_s$, $\dot{M}$ can be re-written as $4 \pi R^2 \dot{R} \rho_s$, where $R$ is the geometric radius of the planetesimal.  
This can be used in Equation (\ref{eqn:mdot_planetesimal}) to represent the growth of the planetesimal's radius as
\begin{equation} \label{eqn:Rdot}
\dot{R} = \frac{1}{4} \frac{f\rho}{\rho_s} v_{\textrm{rel}} \frac{{R^\prime}^2}{R^2}.
\end{equation}
Further understanding the growth depends on the behaviour of $R^\prime$.  
To proceed, we consider two limits: gravitational focusing and Bondi/pebble accretion.  
In the gravitational focusing regime, $R^\prime = R \sqrt{1 + \frac{v_{\textrm{esc}}^2}{v_{\textrm{rel}}^2}}$ (e.g., Equation \ref{eqn:acc_grav_rate}), in which gas only affects the accretion rate through its role in determining $v_{\rm rel}$. 
We first focus on this limit, which we denote as $\dot{R}\vert_G$.


Rewriting the escape speed,  $\dot{R}\vert_G$ in the gravitational focusing limit becomes
\begin{equation} \label{eqn:rgrow_grav}
\dot{R}\vert_G = \frac{1}{4} \frac{f \rho}{\rho_s} v_{\textrm{rel}}+ \frac{2 \pi f \rho G R^2}{3 v_{\textrm{rel}} }.
\end{equation}
The first term in Equation (\ref{eqn:rgrow_grav}) is the contribution from the geometric cross-section.  As gravitational focusing becomes more important, the second term dominates while the first term becomes negligible.  Equation (\ref{eqn:rgrow_grav}) has the form $\dot{R} = AR^2 + B$ with the solution 
\begin{equation} \label{eqn:rsize_grav}
R(\textrm{t})|_\textrm{G} =  v_\textrm{rel}  \sqrt{ \frac{3}{8 \pi G \rho_s}} \textrm{tan} \Bigg [ f \rho \sqrt{ \frac{\pi G}{6 \rho_s}} \textrm{t} + \textrm{arctan} \bigg ( \frac{R_0}{v_\textrm{rel}} \sqrt{\frac{8 \pi G \rho_s}{3} } \bigg ) \Bigg] .
\end{equation}

The planetesimal size reaches a runaway ($R\rightarrow\infty$) when the argument of the tangent approaches $\frac{\pi}{2}$; thus,
\begin{equation} \label{eqn:tgrow_grav}
\textrm{t}\vert_\textrm{G} = \frac{1}{f \rho} \sqrt{\frac{6 \rho_s}{\pi G}} \Bigg [  \frac{\pi}{2} - \textrm{arctan} \bigg ( \frac{R_0}{ v_\textrm{rel} }  \sqrt{\frac{8 \pi G \rho_s}{3} } \bigg) \Bigg], 
\end{equation}
where $\textrm{t}\vert_\textrm{G}$ is the runaway time in the gravitational focusing limit.  

Some important scalings are not immediately clear from this form, so we return to Equation (\ref{eqn:rgrow_grav}), but ignore the geometric term (i.e., assume we are already in the strong gravitational focusing limit).  
 In this case, the approximate solution to $\dot{R}(\textrm{t})\vert_G$ is, 
\begin{equation} \label{eqn:tgrow_grav_simp}
R\textrm{(t)}|_G \approx \frac{3 v_\textrm{rel} R_0}{3  v_\textrm{rel}  - 2 \pi f \rho G R_0 \textrm{t}},
\end{equation}
which has a clear runaway when $t = \frac{3 v_\textrm{rel} }{2 \pi f \rho G R_0} $.  
As planetesimals approach the strong focusing limit, the runaway time is inversely proportional to the initial size of the planetesimal itself.

The second limit we consider is set by the Bondi radius rather than the gravitational focusing radius, $R^\prime = \frac{G m}{v_{\textrm{rel}} ^2}$, which also assumes that gas drag is very effective in promoting the accretion of small solids.  As we saw in the wind tunnel simulations, the efficiency of such accretion is highly dependent on the size of the small solids.  
As such, we assume that particles that are accreted remain within the most efficient size scale (which actually varies as the object increases in size). 
Moreover, the maximum accretion rate seems to occur when the stopping distance is a fraction of the Bondi radius (Table \ref{tbl:bigtable}), so using the full Bondi radius should be understood as a best case scenario.
Using the new expression for $R^\prime$, Equation (\ref{eqn:Rdot}) becomes
\begin{equation} \label{eqn:rrate_bondi}
\dot{R}\vert_B = \frac{4 \pi^2 f \rho \rho_s R^4 G^2}{9  v_{\textrm{rel}} }, 
\end{equation}
where the subscript $B$ denotes the Bondi regime. 
 Again integrating the $\dot{R}$ expression, the planetesimal radius as a function of time is
\begin{equation} \label{eqn:rgrow_bondi}
R(t)\vert_B = \Bigg ( \frac{3  v_{\textrm{rel}}^3 R_0^3}{3 v_{\textrm{rel} }^3 - 4 \pi^2 G^2 f \rho \rho_s R_0^3 \textrm{t}} \Bigg ) ^ {\frac{1}{3}}.
\end{equation}

Runaway accretion occurs as the denominator approaches zero; thus,
\begin{equation} \label{eqn:tgrow_bondi}
t\vert_B = \frac{3  v_{\textrm{rel}}^3}{4 \pi^2 G^2 f \rho \rho_s R_0^3}.
\end{equation}
Note that this timescale depends inversely on $R_0^3$, exhibiting a much stronger dependence on the initial size than the gravitational focusing limit.

\begin{figure}
\centering
\includegraphics[scale=0.66]{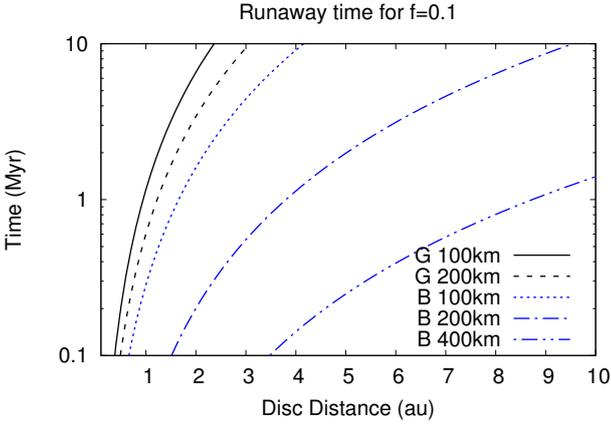}
\caption{Time to reach runaway growth ($R\rightarrow\infty$) for planetesimals with different initial radii ($R=100$, 200, and 400 km).  
The curves show the results assuming growth in the gravitational focusing limit $G$ and the Bondi limit $B$ as a function of distance within the envisaged disc (Equation \ref{eqn:disc}).
All calculations assume that $f=0.1$, which accounts for settling within the disc.  While different disc models will change the results, the curves highlight that runaway growth is very sensitive to the initial planetesimal size.}\label{fig:runaway_times}
\end{figure}

Using the disc model from Equation (\ref{eqn:disc}) and assuming a planetesimal with initial radius $R_0 = 100$ km and $\rho_s = 3 \frac{g}{cm^3}$, we estimate the time required to reach runaway growth in both the gravitational focusing limit and the Bondi accretion limit. We find that runaway timescales for both limits are exceedingly long ($> 1 \times 10^{7}$ Myr) at 10 au, and decrease to $< 3 \times 10^{6}$ at 1 au in both limits.  At 0.1 au, runaway time is only 3800 yr under gravitational focusing, and 950 yr under the Bondi limit.  For all distances, Bondi accretion is roughly four times shorter than accretion from the gravitational focusing limit. 
We assume that significant settling has taken place, with $f=0.1$, leading to more rapid growth \citep{Orm2010}. 
The range  of timescales and the importance of the initial radius are further emphasized in Figure \ref{fig:runaway_times}.  
For the given disc model, rapid growth through small solids (in either limit) is only possible close to the star whenever the initial planetesimal radius is $\lesssim 100$ km.
However, for seed planetesimals that form with sizes well in excess of 100 km, rapid formation is possible out to at least 10 au.  
Again, this result depends on the degree of settling and the actual disc mass.




\subsection{Accretion-driven drift}

One of the assumptions underlying the growth models explored here is that the planetesimal is accreting solids that are on sub-Keplerian orbits.  
This implies that the planetesimal, which is assumed to be on a Keplerian orbit, will preferentially accrete material that has low specific angular momentum compared with the planetesimal's specific angular momentum.  
This will result in a small torque that can cause some degree of migration in the disc, in addition to modest migration introduced by gas drag. 

We first consider planetesimal migration due to gas drag (e.g., Adachi et al.~1976), which leads to a corresponding torque on the planetesimal of
\begin{equation} \label{eqn:dt2} 
\vec{\Gamma_d} = - \frac{1}{2} c_D \rho \pi R^2 v_{\textrm{rel}}^2 r \hat{e_z},
\end{equation}
in which  $c_D$ is the drag coefficient (assumed to be $ \approx 0.5$ for a sphere in the planetesimal limit), $R$ is the planetesimal radius, and $r$ is the stellar separation.  
Here, we assume that the $v_{\rm rel}$ is a headwind.  We further restrict motions to be within the midplane of the disc, which is perpendicular to the $\hat{e_z}$ direction.  
Hereafter, we drop the explicit unit vector, with it being understood that we are exploring radial motions in the midplane and that the relevant torques and angular momenta are in the $\pm \hat{e_z}$ direction.

Given a constant mass, the time-derivative of the orbital angular momentum is, 
\begin{equation} \label{eqn:amderiv}
\frac{d}{dt} (m (G M_{\star} r) ^\frac{1}{2}) |_d = \frac{1}{2} m v_{\textrm{kep}} \dot{r}|_d,
\end{equation}
where $m$ is the planetesimal mass and $v_{\textrm{kep}}$ is the circular Keplerian speed. We assume that the planetesimal remains on a circular orbit throughout its evolution.  
Relating the torque and angular momentum yields
\begin{equation} \label{eqn:tqam2}
\frac{1}{2} m v_{\textrm{kep}} \dot{r}|_d = - \frac{1}{2} c_D \rho \pi R^2 v_{\textrm{rel}}^2 r.
\end{equation}
Solving for the drift rate due to drag gives 
\begin{equation} \label{eqn:rdotd}
\dot{r}\vert_d = - \frac{3}{4} c_D \frac{\rho}{\rho_s} \frac{r}{R} \frac{v_{\textrm{rel}}^2}{v_{\textrm{kep}}},
\end{equation}
where the planetesimal (solid) density is $\rho_s$.  

The drag torque is complemented by an accretion torque, $\Gamma_a$, due to the accretion of  small solids.  In this case, 
\begin{equation} \label{eqn:at}
\Gamma_a = r \dot{m} ( v_\textrm{kep} - v_{\textrm{rel}}  ). 
\end{equation}
Because the mass is no longer constant,
\begin{equation} \label{eqn:atderiv}
\frac{d}{dt} (m (G M_{\star} r) ^\frac{1}{2})|_a = \dot{m} v_{\textrm{kep}} r + \frac{m}{2} v_{\textrm{kep}} \dot{r}|_a .
\end{equation}
Equating this expression with the accretion torque and solving for $\dot{r}\vert_a$ yields
\begin{equation} \label{eqn:rdota}
\dot{r}\vert_a = -2 \frac{\dot{m}}{m} \frac{r}{v_{\textrm{kep}}} {v_\textrm{rel}}.
\end{equation}
This is an accretion-only process. When $v_\textrm{rel} = 0$, there is no migration due to accretion because the specific angular momentum does not change. 

If the planetesimal grows from its Bondi sphere under the pebble accretion paradigm, then
\begin{equation}\label{eqn:bondi_mdot}
\dot{m}=f\rho\pi G^2 M^2 v_{\rm rel}^{-3},
\end{equation}
which means 
\begin{equation}
\dot{r}\vert_a = -2 r \frac{ f\rho\pi G^2 M}{v_{\textrm{kep}} v_{\rm rel}^{2} }.
\end{equation}

In this framework, the $total$ drift rate $\dot{r}$ is the sum of the drift due to drag and accretion, $\dot{r}\vert_d + \dot{r}\vert_a $.  Strictly, the growth and drift are co-dependent. 
For this reason, we solve for $\dot{r}$ numerically using a fourth-order Runge-Kutta integrator, following the mass growth (under Equation \ref{eqn:bondi_mdot}) and the migration rates up until we expect runaway growth to occur.  
Nonetheless, the accretion is a runaway process.  
As such, the overall affect of the accretion torque can be approximated by assuming that the entire planetesimal mass is accreted at the initial planetesimal position, but from a sub-Keplarian solid population.  In this case, the final distance $r_f$ is related to the initial distance $r_i$ by
\begin{equation}\label{eqn:drift_simple}
r_f = r_i \left(1-\frac{v_{\rm rel}}{v_{\rm kep}}\right)^2.
\end{equation}
The results of the full integration, as well as the simple estimate, are given in Table \ref{tbl:angr100}, assuming an initial planetesimal of $R=100$ km and $\rho=3$ g/cc. 
The difference between the simple calculation result and the full integration can be reconciled mainly by the contribution of the gas drag torque, which has increased importance as the time for a runaway increases.
The migration that results from the growth of the planetesimals is less than 1 percent of the initial planetesimal orbital distance, under the current conditions.  
The 10 au planetesimal does not undergo a runaway, which is why it does not follow the trend of a larger inward drift fraction with increasing distance.
Overall, such migration is too small to account for any large-scale changes in the planetesimal distribution, but it may move additional planetesimals into resonances should planets be present in the disc.

\begin{table*}
\centering
\caption{ Inward drift resulting from gas drag and accretion of low specific angular momentum solids for select initial disc radii $r_i$.  
Planetesimals with $R=100$ km and $\rho_s=3$ g/cc are all integrated until they reach runaway growth or, in the case of the 10 au planetesimal, reach a timescale of 10 Myr. 
The change in orbital distance $\Delta r = r_f-r_i$, for final disc radius $r_f$, are all less than about 1\%, with the majority of the inward drift due to the accretion torque.  
The column with $\Delta r_a$ is calculated under the assumptions of Equation (\ref{eqn:drift_simple}).}\label{tbl:angr100}
\smallskip
\begin{tabular}{|>{}l||l|l|l|}
\hline
$r_i$ & Integration Time (yr) & $\frac{\Delta r}{r_i}$ $(\%)$ & $\frac{\Delta r_{\rm a}}{r_i}$ $(\%)$ \\ \hhline{|=||=|=|=|}
    10 & $10\times10^6$ &  -0.15 & -1.2 \\ \hline
    3 & $4.6\times10^6$  & -0.89    & -0.63 \\ \hline
    1 & $2.9\times10^5$ & -0.51 & -0.36   \\ \hline
    0.3 & $14\times10^3$ & -0.27 & -0.20   \\ \hline
    0.1 & 900  & -0.15 & -0.11 \\ \hline
  \end{tabular}
\end{table*}

\section{Discussion and Conclusions}

\subsection{Summarized Results}
In this work, we have investigated the efficiency of the accretion of small pebbles onto planetesimal seeds within protoplanetary discs.  
Our hydrodynamics simulations highlight three basic behaviours, which are captured in Figure \ref{fig:final_accr}.  (1) Nearly every position in the disc for a given planetesimal shows a preferred solid size for accretion, as expected from analytics.  
(2) Solids that are much larger than this preferred size accrete at a rate that is consistent with gravitational focusing alone.  (3) For solid sizes much smaller than the preferred accretion size, the coupling between the gas and the solids is such that the solids remain entrained with the gas as they flow around the planetesimal.   This effect can significantly reduce the planetesimal's effective accretion area to be much smaller than its geometric cross-section. The resulting transition to low accretion rates can be very sensitive to particle size, as well, with accretion efficiency sharply dropping off in some cases.

Centimetre-sized pebbles are most efficiently accreted at 1 au and inwards, while sub-millimetre sizes are preferred at large stellar separations. 
 Chondrule-sized objects in particular are preferentially accreted between 1 au and 3 au (for the given conditions), which overlaps with meteorite parent bodies.  
 In contrast, the low-density, 100 km planetesimal at 10 au shows no preference in the accretion of small solids, with minimally less efficient accretion of micron-sized grains.  
We thus expect small planetesimals that formed at large stellar separations to roughly preserve the size distribution of solids at the time of formation, even after phases of small solid accretion.  
However, as the initial mass of the host planetesimal increases, particles with sizes between 10 $\mu$m and 0.1 mm can become preferentially accreted at this distance, depending on the planetesimal mass.  
This can be seen in the accretion rates of both the high-density 100-km planetesimal as well as the 150-km planetesimal.  


If planetesimals do grow via pebble accretion, then the particle size distribution that can be recovered from an undifferentiated planetesimal will exhibit the following: (1) There will be an optimal pebble size for planetesimal mass and formation location. This optimal pebble size can in principle change as the planetesimal grows in size/mass.  (2) Pebbles much larger than the optimal size will reflect the size distribution of large solids in the nebula.  (3) Solids much smaller than the optimum pebble size will naturally be filtered from planetesimal accretion, truncating the lower end of the size solid distribution in undifferentiated meteorites.  Small grains, e.g., chondritic matrix material \citep{scott_krot2005}, would need to be accreted as aggregates.  


Our simulations and analytic estimates further allow us to explore planetesimal growth timescales.  Even under optimized conditions, runaway growth timescales exceed 10 Myr  at distances greater than 3 au for Bondi accretion, assuming a 100 km planetesimal with an internal density of 3 $\frac{g}{cm^3}$. The runaway timescale for accretion in the Bondi limit (Equation \ref{eqn:tgrow_bondi}) has a $\frac{1}{{R_0}^3}$ dependence, so it is highly sensitive to the size of the initial planetesimal. Larger planetesimals with initial radii of 200 km and 400 km (Fig.~\ref{fig:runaway_times}) reach runaway growth within 3 Myr at up to 5 au and beyond 10 au, respectively.  If some mechanism can produce planetesimals of this size, pebble accretion becomes very efficient.  Subsequent numerical integration of inward drift due to angular momentum transfer shows that planets drifting in $\lesssim 1$ \% of their initial distance.  While this is a small effect under our assumed conditions, it may be able to place planetesimals or embryos into resonances with other objects.

\subsection{Comparison to Other Work}
 
This paper builds upon previous work by \cite{Joh2015}, who investigated planetesimal growth through chondrule accretion. Their work provides evidence for the formation of a distribution of $>$100 km sized planetesimals by the streaming instability, and subsequent growth through the accretion of solids and other planetesimals.  They ran simulations from 2.5 to 25 au using a size distribution of mm-pebbles and smaller, as well as a few with cm-sized pebbles.  They also find that $\sim 1$ mm sized pebbles are size sorted at $\sim$ 3 au, and that the optimally accreted pebble size increases with planetesimal size.  Their 1 au simulations have lower accretion rates than what we recover here, likely owing to the small size of the pebbles.  While sedimentation increases accretion rates at 10 au, growth to planetesimal sizes is more rare than at other distances. Their simulations that were run with cm sized pebbles at 1 au did raise accretion rates.  Our work recovers similar conclusions about the growth of a 100 km planetesimal, and looks specifically at the influence of a range of solid sizes on accretion rates under different conditions using direct simulations of gas dynamics with particle coupling.

Related work by \cite{Bit2015} found that initial planetesimals can reach core masses at all orbital distances.  Their work simulated the growth of planetesimals with larger initial radii ($\sim$600 km to 1700 km) than what we considered in our simulations. However, our analytic timescales for runaway growth under optimum Bondi accretion are in good agreement with runaway growth for planetesimals that already start at these large sizes.  Our simulated planetesimals are smaller than those modeled by \citep{Morb2012} and \citep{Kret2014}; nevertheless, we find reasonable agreement between our accretion rates for a 150 km planetesimal and those of a 200 km planetesimal in the latter.

This work distinguishes itself from previous studies by directly simulating the gas dynamics along with the particle flow and by providing a framework for the analytic estimates for runaway growth times for optimum Bondi-like accretion.   Our simulations explore a wide range of pebble sizes over a wide range of disc conditions.

\subsection{Caveats}

  In optimized conditions, we find that pebble accretion can be very rapid and cause the planetesimal to reach runaway growth well within the time constraint set by gas giants, meteorite chronologies, and observed protoplanetary discs.  We make a number of assumptions about our disc, which can be easily modified for our analytic calculations. 
  Our disc profiles assume outward flaring outlined in Equation \ref{eqn:disc}, which may not be a precise representation of disc structure.  The dust-to-gas ratio of 0.01 ascribed to the midplane in our gas dynamics simulations with particles may be a lower estimate, while larger values due to settling or higher metallicity could enhance accretion.  A high dust-to-gas ratio may further allow accretion of small particles through gas feedback effects.
  The most efficient solid size in our simulations is dependent on our disc model, but the reasonable agreement between our results and general analytic expectations can be used to scale the behaviour.  Moreover, due to the coarse sampling of particle size, the true peak particle size (and accretion rate) may be at an intermediate value.

\subsection{Future Work}

Going forward, these simulations can be refined to be fully 3D rather than relying on the 2D-3D method that we used in this work.  Our simulations do not take into account radial drift rates of pebbles, which could provide an additional size-dependent effect on the accretion rates and could lead to rotational effects on the accreting body.

The OSIRIS-REx (Origins, Spectral Interpretation, Resource Identification, Security - Regolith Explorer) mission is expected to recover $\sim60$ g of regolith from the asteroid Bennu  \citep{lauretta2017}, which may provide clues how material in the asteroid was assembled.   Much more distant missions or ventures in asteroid mining  would be a critical measure of the internal structure, composition, and size distribution of solids in parent bodies.

A.G.H.~and A.C.B.~acknowledge support from an NSERC Discovery Grant, the Canadian Foundation for Innovation, and the British Columbia Knowledge and Development Fund, and The University of British Columbia.  We thank the anonymous referee for comments that improved this manuscript.

%















\label{lastpage}
\end{document}